\documentstyle[12pt]{article}

\font\twlgot =eufm10 scaled \magstep1 \font\egtgot =eufm8
\font\sevgot =eufm7 \font\twlmsb =msbm10 scaled \magstep1
\font\egtmsb =msbm8 \font\sevmsb =msbm7

\newfam\gotfam
\def\pgot{\fam\gotfam\twlgot}
\textfont\gotfam\twlgot \scriptfont\gotfam\egtgot
\scriptscriptfont\gotfam\sevgot
\def\got{\protect\pgot}
\newfam\msbfam
\textfont\msbfam\twlmsb \scriptfont\msbfam\egtmsb
\scriptscriptfont\msbfam\sevmsb
\def\Bbb{\protect\pBbb}
\def\pBbb{\relax\ifmmode\expandafter\Bb\else\typeout{You cann't use
Bbb in text mode}\fi}
\def\Bb #1{{\fam\msbfam\relax#1}}

\newcommand{\gE}{{\got E}}

\newcommand{\gJ}{{\got J}}

\newcommand{\gS}{{\got S}}
\newcommand{\gF}{{\got F}}

\newcommand{\ccG}{{\got g}}
\newcommand{\gA}{{\got A}}
\newcommand{\gL}{{\got L}}

\def\thebibliography#1{\section*{References}\list
  {[\arabic{enumi}]}{\settowidth\labelwidth{#1}\leftmargin\labelwidth
    \advance\leftmargin\labelsep
    \usecounter{enumi}}
    \def\newblock{\hskip .11em plus .33em minus .07em}
    \sloppy\clubpenalty4000\widowpenalty4000
    \sfcode`\.=1000\relax}

\def\op#1{\mathop{\fam0 #1}\limits}

\newcommand{\beq}{\begin{equation}}
\newcommand{\eeq}{\end{equation}}
\newcommand{\ben}{\begin{eqnarray}}
\newcommand{\een}{\end{eqnarray}}
\newcommand{\be}{\begin{eqnarray*}}
\newcommand{\ee}{\end{eqnarray*}}
\newcommand{\bea}{\begin{eqalph}}
\newcommand{\eea}{\end{eqalph}}
\newcommand{\cA}{{\cal A}}

\newcommand{\cP}{{\cal P}}
\newcommand{\cD}{{\cal D}}

\newcommand{\cC}{{\cal C}}
\newcommand{\cL}{{\cal L}}
\newcommand{\cE}{{\cal E}}

\newcommand{\cF}{{\cal F}}

\newcommand{\cS}{{\cal S}}
\newcommand{\cO}{{\cal O}}
\newcommand{\bL}{{\bf L}}

\newcommand{\bE}{{\bf E}}
\newcommand{\bu}{{\bf u}}
\newcommand{\bb}{{\bf b}}
\newcommand{\rL}{{\rm L}}

\newcommand{\al}{\alpha}
\newcommand{\vr}{\varrho}
\newcommand{\bt}{\beta}
\newcommand{\dl}{\delta}
\newcommand{\la}{\lambda}
\newcommand{\La}{\Lambda}
\newcommand{\f}{\phi}
\newcommand{\om}{\omega}

\newcommand{\m}{\mu}

\newcommand{\g}{\gamma}
\newcommand{\G}{\Gamma}
\newcommand{\th}{\theta}

\newcommand{\vt}{\vartheta}

\newcommand{\up}{\upsilon}

\newcommand{\di}{{\rm dim\,}}

\newcommand{\si}{\sigma}
\newcommand{\Si}{\Sigma}
\newcommand{\w}{\wedge}
\newcommand{\wt}{\widetilde}

\newcommand{\ol}{\overline}
\newcommand{\dr}{\partial}
\newcommand{\ar}{\op\longrightarrow}
\newcommand{\ot}{\otimes}
\newcommand{\ap}{\approx}
\newcommand{\ve}{\varepsilon}

\newcommand{\lto}{\leftarrow}
\newcommand{\im}{{\rm Im\,}}
\newcommand{\llr}{\op\longleftarrow}

\newcommand{\rdr}{\stackrel{\leftarrow}{\dr}{}}

\newcounter{theorem}
\newcounter{remark}
\newcounter{proposition}
\newcounter{lemma}
\newcounter{corollary}
\newcounter{definition}

\def\theremark{\arabic{remark}}

\def\thedefinition{\arabic{definition}}

\newenvironment{theo}{\refstepcounter{definition} \medskip
\noindent{\bf Theorem \thedefinition.}}{\medskip}

\newcommand{\mar}[1]{}

\begin{document}
\hbox{}

\begin{center}

{\large\bf CLASSICAL FIELD THEORY. ADVANCED MATHEMATICAL
FORMULATION}
\bigskip

{\sc G.SARDANASHVILY}

{\it Department of Theoretical Physics, Moscow State University,
117234, Moscow, Russia}

\end{center}

{\parindent=0pt

\begin{small}

\bigskip

In contrast with QFT, classical field theory can be formulated in
strict mathematical terms of fibre bundles, graded manifolds and
jet manifolds. Second Noether theorems provide BRST extension of
this classical field theory by means of ghosts and antifields for
the purpose of its quantization.

\bigskip


\end{small}

}

\section{Introduction}

Contemporary QFT is mainly developed as quantization of classical
field models. In contrast with QFT, classical field theory can be
formulated in a strict mathematical way that we present.

Observable classical fields are an electromagnetic field, Dirac
spinor fields and a gravitational field on a world real smooth
manifold. Their dynamic equations are Euler--Lagrange equations
derived from a certain Lagrangian. One also considers classical
non-abelian gauge fields and Higgs fields. Basing on these models,
we develop Lagrangian theory of classical Grassmann-graded (even
and odd) fields on an arbitrary smooth manifold in a very general
setting. Geometry of fibre bundles is known to provide the
adequate mathematical formulation of classical gauge theory and
gravitation theory. Generalizing this formulation, we define even
classical fields as sections of smooth fibre bundles and,
accordingly, develop classical field theory as dynamic theory on
fibre bundles. It is conventionally formulated in terms of jet
manifolds \cite{ald,bau,bry,book,got91,herm,kras,tak1,book09}.

Note that we are in the category of finite-dimensional smooth real
manifolds, which are Hausdorff, second-countable and,
consequently, paracompact. Let $X$ be such a manifold. If
classical fields form a projective $C^\infty(X)$-module of finite
rank, their representation by sections of a fibre bundle follows
from the well-known Serre--Swan theorem extended to non-compact
manifolds \cite{book05}.

Lagrangian theory on fibre bundles is algebraically formulated in
terms of the variational bicomplex of exterior forms on jet
manifolds \cite{ander,bau,jmp,cmp04,olv,tak2,tul,book09}. We are
not concerned with solutions of field equations, but develop
classical field theory as {\it sui generis} prequantum theory that
necessarily involves odd fields. For instance, these are ghosts
and antifields in the second Noether theorem.

There are different descriptions of odd fields in terms of graded
manifolds \cite{cari03,mont06} and supermanifolds
\cite{cia95,franc}. Both graded manifolds and supermanifolds are
described in terms of sheaves of graded commutative algebras
\cite{bart,book00}. However, graded manifolds are characterized by
sheaves on smooth manifolds, while supermanifolds are constructed
by gluing of sheaves on supervector spaces. Treating odd fields on
a smooth manifold $X$, we follow the Serre--Swan theorem
generalized to graded manifolds \cite{jmp05a}. This states that,
if a Grassmann $C^\infty(X)$-algebra is an exterior algebra of
some projective $C^\infty(X)$-module of finite rank, it is
isomorphic to the algebra of graded functions on a graded manifold
whose body is $X$.

Lagrangian theory on fibre bundles is generalized to Lagrangian
theory of even and odd variables on graded manifolds in terms of
the Grassmann-graded variational bicomplex
\cite{barn,jmp05,jmp05a,cmp04,book09}. Theorem \ref{t1} on
cohomology of the variational bicomplex results in a solution of
the global inverse problem of the calculus of variations (Theorem
\ref{c1}), the first variational formula (Theorem \ref{t2}) and to
the first Noether theorem in a very general setting of
supersymmetries depending on higher-order derivatives of fields
(Theorem \ref{t3}).

Quantization of Lagrangian field theory essentially depends on its
degeneracy, characterized by non-trivial Noether  and higher-stage
Noether identities, and implies its BRST extension  by means of
the corresponding ghosts and antifields
\cite{barn,lmp08,fust,gom}.

Any Euler--Lagrange operator satisfies Noether identities
(henceforth NI) which are separated into the trivial and
non-trivial ones. These NI obey first-stage NI, which in turn are
subject to the second-stage NI, and so on. However, there is a
problem how to select trivial and non-trivial higher-stage NI. We
follow the general notion of NI of a differential operator
\cite{oper}. They are represented by one-cycles of a certain chain
complex. Its boundaries are trivial NI, and non-trivial NI modulo
the trivial ones are given by first homology of this complex. To
describe $(k+1)$-stage NI, let us assume that non-trivial
$k$-stage NI are generated by a projective $C^\infty(X)$-module
$\cC_{(k)}$ of finite rank and that a certain homology condition
holds \cite{jmp05a,lmp08,book09}. In this case, $(k+1)$-stage NI
are represented by $(k+2)$-cycles of some chain complex of modules
of antifields isomorphic to $\cC_{(i)}$, $i\leq k$, by virtue of
the Serre--Swan theorem. Accordingly, trivial $(k+1)$-stage NI are
defined as its boundaries. Iterating the arguments, we come to the
exact Koszul--Tate (henceforth KT) complex (\ref{v94}) with the
boundary KT operator (\ref{v92}) whose nilpotentness is equivalent
to all non-trivial NI (Theorem \ref{t4})
\cite{jmp05a,lmp08,book09}.

The inverse second Noether theorem (Theorem \ref{t5}) associates
to the KT complex the cochain sequence (\ref{w36}) with the ascent
operator $\bu$ (\ref{w108}), called the gauge operator. Its
components (\ref{w33}) and (\ref{w38}) are non-trivial gauge and
higher-stage gauge symmetries of Lagrangian theory. They obey the
gauge symmetry conditions (\ref{w19}) and (\ref{w34}). The gauge
operator unlike the KT one is not nilpotent, unless gauge
symmetries are abelian. Therefore, in contrast with NI, an
intrinsic definition of non-trivial gauge and higher-stage gauge
symmetries meets difficulties. Defined by the gauge operator,
gauge and higher-stage gauge symmetries are indexed by ghosts, not
gauge parameters. Herewith,  $k$-stage gauge symmetries act on
$(k-1)$-stage ghosts.

Gauge symmetries fail to form an algebra in general
\cite{fulp02,jmp08,gom}. We say that gauge and higher-stage gauge
symmetries are algebraically closed if the gauge operator $\bu$
(\ref{w108}) admits the nilpotent BRST extension $\bb$
(\ref{w109}) where $k$-stage gauge symmetries are extended to
$k$-stage BRST transformations acting both on $(k-1)$-stage and
$k$-stage ghosts \cite{jmp08,book09}. The BRST operator
(\ref{w109}) brings the cochain sequence (\ref{w36}) into the BRST
complex.

The KT and BRST complexes provide an above mentioned BRST
extension of original Lagrangian field theory. This extension
exemplifies so called field-antifield theory whose Lagrangians are
required to satisfy the particular condition (\ref{w44}) called
the classical master equation. We show that an original Lagrangian
is extended to a proper solution of the master equation if the
gauge operator (\ref{w108}) admits a nilpotent BRST extension
(Theorem \ref{t6}) \cite{lmp08,book09}.

Given the BRST operator (\ref{w109}), a desired proper solution of
the master equation is constructed by the formula (\ref{w133}).
This construction completes the BRST extension of original
Lagrangian theory to the prequantum one, quantized in terms of
functional integrals.

The basic field models, including gauge theory, gravitation theory
and spinor fields, are briefly considered.

\section{Jet manifolds}

Jet formalism \cite{ander,book,kol,ijgmmp07,sau,tak2} provides the
conventional language of theory of differential equations and
Lagrangian theory on fibre bundles
\cite{bau,bry,book,got91,herm,kras,tak1,book09}.

Given a smooth fibre bundle $Y\to X$, a $k$-order jet $j^k_xs$ at
a point $x\in X$ is defined as an equivalence class of sections
$s$ of $Y\to X$ identified by $k+1$ terms of their Taylor series
at $x$. A key point is that a set $J^kY$ of $k$-order jets is a
finite-dimensional smooth manifold coordinated by $(x^\la,y^i,
y^i_\la, \ldots, y^i_{\la_k\ldots\la_1})$, where $(x^\la,y^i)$ are
bundle coordinates on $Y\to X$ and $y^i_{\la_r\ldots\la_1}$ are
coordinates of derivatives, i.e., $y^i_{\la_r\ldots\la_1}\circ
s=\dr_{\la_r}\cdots\dr_{\la_1}s(x)$. Accordingly, the infinite
order jets are defined as equivalence classes of sections of a
fibre bundle $Y\to X$ identified by their Taylor series. Infinite
order jets form a paracompact Fr\'echet (not smooth) manifold
$J^\infty Y$. It coincides with the projective limit of the
inverse system of finite order jet manifolds
\mar{5.10}\beq
X\op\longleftarrow Y\op\longleftarrow J^1Y \longleftarrow \cdots
J^{r-1}Y \op\longleftarrow J^rY\longleftarrow\cdots. \label{5.10}
\eeq

The main advantage of jet formalism is that it enables one to deal
with finite-dimensional jet manifolds instead of
infinite-dimensional spaces of fields. In the framework of jet
formalism, a $k$-order differential equation on a fibre bundle
$Y\to X$ is defined as a closed subbundle $\gE$ of the jet bundle
$J^kY\to X$. Its solution is a section $s$ of $Y\to X$ whose jet
prolongation $J^ks$ lives in $\gE$. A $k$-order differential
operator on $Y\to X$ is defined as a morphism of the jet bundle
$J^kY\to X$ to some vector bundle $E\to X$. However, the kernel of
a differential operator need not be a differential equation.

Jet manifolds provide the language of modern differential
geometry. Due to the canonical bundle monomorphism $J^1Y\to
T^*X\ot TY$ over $Y$, any connection $\G$ on a fibre bundle $Y\to
X$ is represented by a global section
\be
\G=dx^\la\ot(\dr_\al +\G^i_\la(x^\m,y^j)\dr_i)
\ee
of the jet bundle $J^1Y\to Y$. Accordingly, we have the $T^*X\ot
VY$-valued first order differential operator
\be
D= (y^i_\al-\G^i_\la)dx^\la\ot\dr_i
\ee
on $Y$. It is called the covariant differential.

Note that there are different notions of jets. Jets of sections
are particular jets of maps \cite{kol} and jets of submanifolds
\cite{book,kras,book09}. Let us mention jets of modules over
commutative rings \cite{kras,book00} and graded commutative rings
\cite{book05}, and of modules over algebras of operadic type
\cite{niep}. Jets of modules over a noncommutative ring however
fail to be defined.

\section{Lagrangian theory of even fields}

We formulate Lagrangian theory on fibre bundles in algebraic terms
of the variational bicomplex
\cite{ander,jmp,cmp04,ijgmmp07,tak2,tul,book09}.

The inverse system (\ref{5.10}) of jet manifolds yields the direct
system
\mar{5.7}\beq
\cO^*X\op\longrightarrow \cO^*Y \op\longrightarrow \cO_1^*Y
\ar\cdots \cO^*_{r-1}Y \op\longrightarrow
 \cO_r^*Y \longrightarrow\cdots  \label{5.7}
\eeq
of differential graded algebras (henceforth DGAs) $\cO_r^*Y$ of
exterior forms on jet manifolds $J^rY$. Its direct limit is the
DGA $\cO_\infty^*Y$ of all exterior forms on finite order jet
manifolds. This DGA is locally generated by horizontal forms
$dx^\la$ and contact forms $\th^i_\La=dy^i_\La
-y^i_{\la+\La}dx^\la$, where $\La=(\la_k...\la_1)$ denotes a
symmetric multi-index, and $\la+\La=(\la\la_k...\la_1)$. There is
the canonical decomposition of $\cO^*_\infty Y$ into the modules
$\cO^{k,m}_\infty Y$ of $k$-contact and $m$-horizontal forms
($m\leq n=\di X$). Accordingly, the exterior differential on
$\cO_\infty^*Y$ falls into the sum $d=d_V+d_H$ of the vertical
differential $d_V:\cO^{k,*}_\infty Y\to \cO^{k+1,*}_\infty Y$ and
the total one $d_H:\cO^{*,m}_\infty Y\to \cO^{*,m+1}_\infty Y$.
One also introduces the projector $\vr$ on $\cO^{>0,n}_\infty Y$
such that $\vr\circ d_H=0$ and the variational operator
$\dl=\vr\circ d$ on $\cO^{*,n}_\infty Y$ such that $\dl\circ
d_H=0$, $\dl\circ\dl=0$. All these operators split the DGA
$\cO^*_\infty Y$ into the variational bicomplex. We consider its
subcomplexes
\mar{b317,f3}\ben
&& 0\to\Bbb R\to \cO^0_\infty Y\ar^{d_H}\cO^{0,1}_\infty Y\cdots
\op\longrightarrow^{d_H} \cO^{0,n}_\infty Y \op\longrightarrow^\dl
\bE_1 \op\longrightarrow^\dl \bE_2 \ar \cdots, \label{b317}\\
&& 0\to \cO^{1,0}_\infty Y\ar^{d_H}\cO^{1,1}_\infty Y\cdots
\op\longrightarrow^{d_H} \cO^{1,n}_\infty Y \op\longrightarrow^\vr
\bE_1\to 0, \qquad \bE_k=\vr(\cO^{k,n}_\infty Y). \label{f3}
\een
Their elements $L\in \cO^{0,n}_\infty Y$ and $\dl L\in \bE_1$ are
finite order Lagrangians on a fibre bundle $Y\to X$ and their
Euler--Lagrange operators.

The algebraic Poincar\'e lemma \cite{olv,tul} states that the
variational bicomplex $\cO^*_\infty Y$ is locally exact. In order
to obtain its cohomology, one therefore can use the abstract de
Rham theorem on sheaf cohomology \cite{hir} and the fact that $Y$
is a strong deformation retract of $J^\infty Y$, i.e., sheaf
cohomology of $J^\infty Y$ equals that of $Y$ \cite{ander,jmp}. A
problem is that the paracompact space $J^\infty Y$ admits the
partition of unity by functions which do not belong to
$\cO^0_\infty Y$. Therefore, one considers the variational
bicomplex ${\cal Q}^*_\infty Y\supset \cO^*_\infty Y$ whose
elements are locally exterior forms on finite order jet manifolds,
and obtains its cohomology \cite{tak2,ijmms}. Afterwards, the
$d_H$- and $\dl$-cohomology of $\cO^*_\infty Y$ is proved to be
isomorphic to that of ${\cal Q}^*_\infty Y$
\cite{lmp,jmp,ijmms,book09}. In particular, cohomology of the
variational complex (\ref{b317}) equals the de Rham cohomology of
$Y$, while the complex (\ref{f3}) is exact.

The exactness of the complex (\ref{f3}) at the last term states
the global first variational formula which, firstly, shows that an
Euler--Lagrange operator $\dl L$ is really a variational operator
of the calculus of variations and, secondly, leads to the first
Noether theorem. Cohomology of the variational complex
(\ref{b317}) at the term $\cO^{0,n}_\infty Y$ provides a solution
of the global inverse problem of the calculus of variations on
fibre bundles. It is the cohomology of variationally trivial
Lagrangians which are locally $d_H$-exact.

\section{Odd fields}

The algebraic formulation of Lagrangian theory of even fields in
terms of the variational bicomplex is generalized to odd fields
\cite{barn,jmp05,jmp05a,cmp04,book09}.

Namely, let a bundle $Y\to X$ of classical fields be a vector
bundle. Then all jet bundles $J^kY\to X$ are also vector bundles.
Let us consider a subalgebra $P^*_\infty Y\subset \cO^*_\infty Y$
of exterior forms whose coefficients are polynomial in fibre
coordinates $y^i$, $y^i_\La$ on these bundles. In particular, the
commutative ring $P^0_\infty Y$ consists of polynomials of
coordinates $y^i$, $y^i_\La$ with coefficients in the ring
$C^\infty(X)$. One can associate to such a polynomial a section of
the symmetric tensor product $\op\vee^m (J^kY)^*$ of the dual of
the jet bundle $J^kY\to X$, and {\it vice versa}. Moreover, any
element of $P^*_\infty Y$ is an element of the
Chevalley--Eilenberg differential calculus over $P^0_\infty Y$.
This construction is extended to the case of odd fields.

In accordance with the Serre--Swan theorem \cite{book05}, if a
Grassmann $C^\infty(X)$-algebra is the exterior algebra of some
projective $C^\infty(X)$-module of finite rank, it is isomorphic
to the algebra of graded functions on a graded manifold
$(X,\cA_F)$ whose a body is $X$ and whose structure ring $\cA_F$
of graded functions consists of sections of the exterior bundle
\be
\w F^*=\Bbb R\op\oplus F^*\op\oplus\op\w^2 F^*\op\oplus\cdots,
\ee
where $F^*$ is the dual of some vector bundle $F\to X$. Then the
Grassmann-graded Chevalley--Eilenberg differential calculus
\be
0\to \Bbb R\to \cA_F \ar^d \cS^1[F;X]\ar^d\cdots
\cS^k[F;X]\ar^d\cdots
\ee
over $\cA_F$ can be constructed. One can think of its elements as
being graded differential forms on $X$. In particular, there is a
monomorphism $\cO^*X\to \cS^*[F;X]$. Following suit of an even DGA
$P^*_\infty Y$, let us consider simple graded manifolds
$(X,\cA_{J^rF})$ modelled over the vector bundles $J^rF\to X$. We
have the direct system of corresponding DGAs
\be
 \cS^*[F;X]\ar
\cS^*[J^1F;X]\ar\cdots \cS^*[J^rF;X]\ar\cdots,
\ee
whose direct limit $\cS^*_\infty[F;X]$ is the Grassmann
counterpart of an even DGA $P^*_\infty Y$.

The total algebra of even and odd fields is the graded exterior
product
\mar{f5}\beq
\cP^*_\infty[F;Y]=P^*_\infty Y\op\w_{\cO^*X} \cS^*_\infty[F;X]
\label{f5}
\eeq
of the DGAs $P^*_\infty Y$ and $\cS^*_\infty[F;X]$ over their
common subalgebra $\cO^*X$ \cite{jmp05,cmp04,book09}. In
particular, $\cP^0_\infty[F;Y]$ is a graded commutative
$C^\infty(X)$-ring whose even and odd generating elements are
sections of $Y\to X$ and $F\to X$, respectively. Let $(x^\la,y^i,
y^i_\La)$ be bundle coordinates on jet bundles $J^kY\to X$ and
$(x^\la, c^a, c^a_\La)$ those on $J^rF\to X$. For simplicity, let
these symbols also stand for local sections $s$ of these bundles
such that $s^i_\La(x)=y^i_\La$ and $s^a_\La(x)=c^a_\La$. Then the
DGA $\cP^*_\infty[F;Y]$ (\ref{f5}) is locally generated by
elements $(y^i,y^i_\La,c^a,c^a_\La, dx^\la,dy^i,dy^i_\La, dc^a,
dc^a_\La)$. By analogy with $(y^i,y^i_\La)$, one can think of odd
generating elements $(c^a,c^a_\La)$ as being (local) odd fields
and their jets.

In a general setting, if $Y\to X$ is not a vector bundle, we
consider graded manifolds $(J^rY,\gA_{F_r})$ whose bodies are jet
manifolds $J^rY$, and $F_r=J^rY\times J^rF$ is the pull-back onto
$J^rY$ of the jet bundle $J^rF\to X$ \cite{jmp05a,lmp08,book09}.
As a result, we obtain the direct system of DGAs
\mar{v6}\beq
\cS^*[Y\times F;Y]\ar \cS^*[F_1;J^1Y]\ar\cdots
\cS^*[F_r;J^rY]\ar\cdots. \label{v6}
\eeq
Its direct limit $\cS^*_\infty[F;Y]$ is a differential calculus
over the ring $\cS^0_\infty[F;Y]$ of graded functions. The
monomorphisms $\cO^*_rY\to \cS^*[F_r;J^rY]$ yield a monomorphism
of the direct system (\ref{5.7}) to that (\ref{v6}) and,
consequently, the monomorphism $\cO^*_\infty Y\to
\cS^*_\infty[F;Y]$ of their direct limits. Moreover,
$\cS^*_\infty[F;Y]$ is a $\cO^0_\infty Y$-algebra. It contains the
$C^\infty(X)$-subalgebra $\cP^*_\infty[F;Y]$ if a fibre bundle
$Y\to X$ is affine. The $\cO^0_\infty Y$-algebra
$\cS^*_\infty[F;Y]$ is locally generated by elements $(c^a,
c^a_\La,dx^\la,dy^i,dy^i_\La, dc^a, dc^a_\La)$ with coefficient
functions depending on coordinates $(x^\la, y^i,y^i_\La)$. One
calls $(y^i,c^a)$ the local basis for the DGA $\cS^*_\infty[F;Y]$.
We further  use the collective symbol $s^A$ for its elements.
Accordingly, $s^A_\La$ denote jets of $s^A$, $\th^A_\La=ds^A_\La-
s^A_{\la+\La}dx^\la$ are contact forms, and $\dr_A^\La$ are graded
derivations of the $\Bbb R$-ring $\cS^0_\infty[F;Y]$ such that
$\dr_{A'}^{\La'}\rfloor ds^A_\La=\dl_{A'}^A\dl_\La^{\La'}$. The
symbol $[A]=[s^A]=[s^A_\La]$ stands for the Grassmann parity.

The DGA $\cS^*_\infty[F;Y]\supset\cO^*_\infty Y$ is split into the
variational bicomplex which describes Lagrangian theory of even
and odd fields.

\section{Lagrangian theory of even and odd fields}

There is the canonical decomposition of the DGA
$\cS^*_\infty[F;Y]$  into the modules $\cS^{k,m}_\infty[F;Y]$ of
$k$-contact and $m$-horizontal graded forms. Accordingly, the
graded exterior differential on $\cS^*_\infty[F;Y]$ falls into the
sum $d=d_V+d_H$ of the vertical differential $d_V$ and the total
differential
\be
&& d_H(\f)= dx^\la\w d_\la(\f),  \qquad d_\la = \dr_\la +
\op\sum_{0\leq|\La|} s^A_{\la+\La}\dr_A^\La, \qquad \f\in
\cS^*_\infty[F;Y],\\
&& d_H\circ h_0= h_0\circ d, \qquad h_0: \cS^*_\infty[F;Y]\to
\cS^{0,*}_\infty[F;Y].
\ee
We also have the graded projection endomorphism $\vr$ of
$\cS^{<0,n}_\infty[F;Y]$ such that $\vr\circ d_H=0$ and the graded
variational operator $\dl=\vr\circ d$ such that $\dl\circ d_H=0$,
$\dl\circ\dl=0$. With these operators the DGA $\cS^*_\infty[F;Y]$
is split into the Grassmann-graded variational bicomplex. It
contains the subcomplexes
\mar{g111,2}\ben
&& 0\to \Bbb R\ar \cS^0_\infty[F;Y]\ar^{d_H}\cS^{0,1}_\infty[F;Y]
\cdots \ar^{d_H} \cS^{0,n}_\infty[F;Y]\ar^\dl \bE_1
=\vr(\cS^{1,n}_\infty[F;Y]), \label{g111} \\
&& 0\to \cS^{1,0}_\infty[F;Y]\ar^{d_H} \cS^{1,1}_\infty[F;Y]\cdots
\ar^{d_H}\cS^{1,n}_\infty[F;Y]\ar^\vr \bE_1\to 0. \label{g112}
\een
One can think of their even elements
\mar{0709,'}\ben
&& L=\cL\om\in \cS^{0,n}_\infty[F;Y], \qquad \om=dx^1\w\cdots \w
dx^n,
\label{0709}\\
&& \dl L= \th^A\w \cE_A\om=\op\sum_{0\leq|\La|}
 (-1)^{|\La|}\th^A\w d_\La (\dr^\La_A L)\om\in \bE_1 \label{0709'}
\een
as being a graded Lagrangian and its Euler--Lagrange operator,
respectively.

The algebraic Poincar\'e lemma states that the complexes
(\ref{g111}) and (\ref{g112}) are locally exact at all the terms,
except $\Bbb R$ \cite{barn,cmp04}. Then one can obtain cohomology
of these complexes in the same manner as that of the complexes
(\ref{b317}) and (\ref{f3}) \cite{ijgmmp07,book09}.

\begin{theo} \label{t1} \mar{t1}
Cohomology of the  variational complex (\ref{g111}) equals the de
Rham cohomology $H^*(Y)$ of $Y$. The complex (\ref{g112}) is
exact.
\end{theo}

Cohomology of the complex (\ref{g111}) at the term
$\cS^{1,n}_\infty[F;Y]$ provides the following solution of the
global inverse problem of the calculus of variation for graded
Lagrangians.

\begin{theo} \label{c1} \mar{c1}
A $\dl$-closed (i.e., variationally trivial) graded density reads
$L_0=h_0\psi + d_H\xi$, $\xi\in \cS^{0,n-1}_\infty[F;Y]$, where
$\psi$ is a non-exact $n$-form on $Y$. In particular, a
$\dl$-closed odd density is $d_H$-exact.
\end{theo}

Exactness of the complex (\ref{g112}) at the last term implies
that any Lagrangian $L$ admits the decomposition
\mar{g99}\beq
dL=\dl L - d_H\Xi,
\qquad \Xi\in \cS^{1,n-1}_\infty[F;Y], \label{g99}\\
\eeq
where $L+\Xi$ is a Lepagean equivalent of $L$. This decomposition
leads to the first variational formula (Theorem \ref{t2}) and the
first Noether theorem (Theorem \ref{t3}).

\section{The first Noether theorem}

In order to treat symmetries of Lagrangian field theory described
by the DGA $\cS^*_\infty[F;Y]$ are defined as contact graded
derivations of the $\Bbb R$-ring $\cS^0_\infty[F;Y]$
\cite{jmp05,cmp04,book09}. Its graded derivation $\vt$ is called
contact if the Lie derivative $\bL_\vt$ of the DGA
$\cS^*_\infty[F;Y]$ preserves the ideal of contact graded forms.
Contact graded derivations take the form
\mar{g105}\beq
\vt=\vt_H+\vt_V=\vt^\la d_\la + (\up^A\dr_A +\op\sum_{|\La|>0}
d_\La\up^A\dr_A^\La), \qquad \up^A=\vt^A-s^a_\m\vt^\m,
\label{g105}
\eeq
where $\vt^\la$, $\vt^A$ are local graded functions.

\begin{theo}  \label{t2} \mar{t2}
It follows from the decomposition (\ref{g99}) that the Lie
derivative $\bL_\vt L$ of a Lagrangian $L$ (\ref{0709}) with
respect to a graded derivation $\vt$ (\ref{g105}) fulfills the
first variational formula
\mar{g107}\beq
\bL_\vt L= \vt_V\rfloor\dl L +d_H(h_0(\vt\rfloor \Xi_L)) + d_V
(\vt_H\rfloor\om)\cL. \label{g107}
\eeq
\end{theo}

In particular, if a vertical graded derivation $\vt$ is treated as
an infinitesimal variation of dynamic variables, then the first
variational formula (\ref{g107}) shows that the Euler--Lagrange
equations $\dl L=0$ are variational equations.

A contact graded derivation $\vt$ (\ref{g105}) is called a
variational symmetry of a Lagrangian $L$ if the Lie derivative
$\bL_\vt L$ of $L$ is $d_H$-exact. One can show that $\vt$ is a
variational symmetry iff its vertical part $\up_V$ (\ref{g105}) is
well. Therefore, we further restrict our consideration to vertical
contact graded derivations
\mar{f6}\beq
\vt=(\up^A\dr_A +\op\sum_{|\La|>0} d_\La\up^A\dr_A^\La).
\label{f6}
\eeq
A glance at the expression (\ref{f6}) shows that such a derivation
is an infinite jet prolongation of its first summand
$\up=\up^A\dr_A$, called the generalized vector field.
Substituting $\vt$ (\ref{f6}) into the first variational formula
(\ref{g107}), we come to the first Noether theorem.

\begin{theo} \label{t3} \mar{t3}
If $\vt$ (\ref{f6}) is a variational symmetry of a Lagrangian $L$
(\ref{0709}) (i.e., $\bL_\up L=d_H\si$, $\si\in
\cS^{0,n-1}_\infty$), the weak conservation law
\be
0\ap d_H(h_0(\vt\rfloor\Xi_L)-\si)
\ee
the Noether current $\gJ_\vt=h_0(\vt\rfloor\Xi_L)$ holds on the
shell $\dl L=0$.
\end{theo}

A vertical graded derivation $\vt$ (\ref{f6}) is called nilpotent
if $\bL_\vt(\bL_\vt\f)=0$ for any horizontal graded form $\f\in
\cS^{0,*}_\infty[F;Y]$. An even graded derivation is never
nilpotent.

For the sake of simplicity, the common symbol further stands for a
generalized vector field $\up$, the contact graded derivation
$\vt$ (\ref{f6}) determined by $\up$  and the Lie derivative
$\bL_\vt$. We agree to call all these operators a graded
derivation of the DGA $\cS^*_\infty[F;Y]$.

\section{The KT complex of Noether identities}

Any Euler--Lagrange operator (\ref{0709'}) obeys NI which are
separated into trivial and non-trivial ones. Trivial NI are
defined as boundaries of a certain chain complex \cite{jmp05a}.
Lagrangian theory is called degenerate if there exists non-trivial
NI. They satisfy first-stage NI and so on. Thus, there is a
hierarchy of reducible NI. Degenerate Lagrangian theory is said to
be $N$-stage reducible if there exists non-trivial $N$-stage NI,
but all $(N+1)$-stage NI are trivial. Under certain conditions,
one can associate to degenerate Lagrangian theory the exact KT
complex whose boundary operator provides all non-trivial NI
(Theorem \ref{t4}) \cite{jmp05a,lmp08,book09}. This complex is an
extension of the original DGA $\cS^*_\infty[F;Y]$ by means of
antifields whose spaces are density-dual to the modules of
non-trivial NI.

Let us introduce the following notation. The density dual of a
vector bundle $E\to X$ is $\ol E^*=E^*\ot\op\w^n T^*X$. Given
vector bundles $E\to X$ and $V\to X$, let $\cS^*_\infty[V\times
F;Y\times E]$ be the extension of the DGA $\cS^*_\infty[F;Y]$
whose additional even and odd generators are sections of $E\to X$
and $V\to X$, respectively. We consider its subalgebra
$\cP^*_\infty[V,F;Y,E]$ with coefficients polynomial in these new
generators. Let us also assume that the vertical tangent bundle
$VY$ of $Y$ admits the splitting $VY=Y\times W$, where $W\to X$ is
a vector bundle. In this case, there no fibre bundle under
consideration whose transition functions vanish on the shell $\dl
L=0$. Let $\ol Y^*$ denote the density-dual of $W$ in this
splitting.

Let $L$ be a Lagrangian (\ref{0709}) and $\dl L$ its
Euler--Lagrange operator (\ref{0709'}). In order to describe NI
which $\dl L$ satisfies, let us enlarge the DGA
$\cS^*_\infty[F;Y]$ to the DGA $\cP^*_\infty[\ol Y^*,F;Y,\ol F^*]$
with the local basis $\{s^A, \ol s_A\}$, $[\ol s_A]=([A]+1){\rm
mod}\,2$. Its elements $\ol s_A$ are called antifields of
antifield number Ant$[\ol s_A]= 1$ \cite{barn,gom}. The DGA
$\cP^*_\infty[\ol Y^*,F;Y,\ol F^*]$ is endowed with the nilpotent
right graded derivation $\ol\dl=\rdr^A \cE_A$. We have the chain
complex
\mar{v042}\beq
0\lto \im\ol\dl \llr^{\ol\dl} \cP^{0,n}_\infty[\ol Y^*;F;Y;\ol
F^*]_1 \llr^{\ol\dl} \cP^{0,n}_\infty[\ol Y^*;F;Y;\ol F^*]_2
\label{v042}
\eeq
of graded densities of antifield number $\leq 2$. Its one-cycles
define the above mentioned NI, which are trivial iff cycles are
boundaries. Accordingly, elements of the first homology
$H_1(\ol\dl)$ of the complex (\ref{v042}) correspond to
non-trivial NI modulo the trivial ones
\cite{jmp05a,lmp08,oper,book09}. We assume that $H_1(\ol \dl)$ is
finitely generated. Namely, there exists a projective
Grassmann-graded $C^\infty(X)$-module $\cC_{(0)}\subset H_1(\ol
\dl)$ of finite rank with a local basis $\{\Delta_r\}$ such that
all non-trivial NI result from the NI
\mar{v64}\beq
\ol\dl\Delta_r= \op\sum_{0\leq|\La|} \Delta_r^{A,\La} d_\La
\cE_A=0. \label{v64}
\eeq

The NI (\ref{v64}) need not be independent, but obey first-stage
NI described as follows. By virtue of the Serre--Swan theorem, the
module $\cC_{(0)}$ is isomorphic to a module of sections of the
product $\ol V^*\times \ol E^*$, where $\ol V^*$ and $\ol E^*$ are
the density-duals of some vector bundles $V\to X$ and $E\to X$.
Let us enlarge the DGA $\cP^*_\infty[\ol Y^*,F;Y,\ol F^*]$ to the
DGA $\cP^*_\infty[\ol E^*\times \ol Y^*,F;Y,\ol F^*\times\ol V^*]$
possessing the local basis $\{s^A,\ol s_A, \ol c_r\}$ of Grassmann
parity $[\ol c_r]=([\Delta_r]+1){\rm mod}\,2$ and antifield number
${\rm Ant}[\ol c_r]=2$. This DGA is provided with the nilpotent
right graded derivation $\dl_0=\ol\dl + \rdr^r\Delta_r$ such that
its nilpotency condition is equivalent to the NI (\ref{v64}). Then
we have the chain complex
\mar{v66}\ben
&&0\lto \im\ol\dl \llr^{\ol\dl} \cP^{0,n}_\infty[\ol Y^*,F;Y,\ol
F^*]_1\llr^{\dl_0}
\cP^{0,n}_\infty[\ol E^*\times \ol Y^*,F;Y,\ol F^*\times\ol V^*]_2 \label{v66}\\
&& \qquad \llr^{\dl_0} \cP^{0,n}_\infty[\ol E^*\times\ol
Y^*,F;Y,\ol F^*\times\ol V^*]_3 \nonumber
\een
of graded densities of antifield number $\leq 3$. It has the
trivial homology $H_0(\ol\dl_0)$ and $H_1(\ol\dl_0)$. The
two-cycles of this complex define the above mentioned first-stage
NI. They are trivial if cycles are boundaries, but the converse
need not be true, unless a certain homology condition holds
\cite{jmp05a,lmp08,oper,book09}. If the complex (\ref{v66}) obeys
this condition, elements of its second homology $H_2(\ol\dl_0)$
define non-trivial first-stage NI modulo the trivial ones. Let us
assume that $H_2(\dl_0)$ is finitely generated. Namely, there
exists a projective Grassmann-graded $C^\infty(X)$-module
$\cC_{(1)}\subset H_2(\dl_0)$ of finite rank with a local basis
$\{\Delta_{r_1}\}$ such that all non-trivial first-stage NI follow
from the equalities
\mar{v82}\beq
\op\sum_{0\leq|\La|} \Delta_{r_1}^{r,\La} d_\La \Delta_r +\ol\dl
h_{r_1} =0. \label{v82}
\eeq

The first-stage NI (\ref{v82}) need not be independent, but
satisfy the second-stage ones, and so on. Iterating the arguments,
we come to the following \cite{jmp05a,lmp08,book09}.

\begin{theo} \label{t4} \mar{t4} One can associate to
degenerate $N$-stage reducible Lagrangian theory the exact KT
complex (\ref{v94}) with the boundary operator (\ref{v92}) whose
nilpotency property restarts all NI and higher-stage NI
(\ref{v64}) and (\ref{v93}) if these identities are finitely
generated and iff this complex obeys the homology regularity
condition.
\end{theo}

Namely, there are vector bundles $V_1,\ldots, V_N, E_1, \ldots,
E_N$ over $X$ and the DGA
\mar{v91}\beq
\ol\cP^*_\infty\{N\}=\cP^*_\infty[\ol E^*_N\times\cdots\times\ol
E^*_1\times\ol E^*\times\ol Y^*,F;Y,\ol F^*\times\ol V^*\times\ol
V^*_1\times\cdots\times\ol V_N^*] \label{v91}
\eeq
with a local basis $\{s^A,\ol s_A, \ol c_r, \ol c_{r_1}, \ldots,
\ol c_{r_N}\}$ of antifield number Ant$[\ol c_{r_k}]=k+2$. Let the
indexes $k=-1,0$ further stand for $\ol s_A$ and $\ol c_r$,
respectively. The DGA $\ol\cP^*_\infty\{N\}$ (\ref{v91}) is
provided with the nilpotent right graded derivation
\mar{v92}\ben
&&\dl_N=\rdr^A\cE_A +
\op\sum_{0\leq|\La|}\rdr^r\Delta_r^{A,\La}\ol s_{\La A} +
\op\sum_{1\leq k\leq N}\rdr^{r_k} \Delta_{r_k},
\label{v92}\\
&& \Delta_{r_k}= \op\sum_{0\leq|\La|}
\Delta_{r_k}^{r_{k-1},\La}\ol c_{\La r_{k-1}} + \op\sum_{0\leq
|\Si|, |\Xi|}(h_{r_k}^{(r_{k-2},\Si)(A,\Xi)}\ol c_{\Si r_{k-2}}\ol
s_{\Xi A}+...), \nonumber
\een
of antifield number -1. It is called the KT differential. With
$\dl_N$, we have the exact chain complex
\mar{v94}\ben
&&0\lto \im \ol\dl \llr^{\ol\dl} \cP^{0,n}_\infty[\ol Y^*,F;Y,\ol
F^*]_1\llr^{\dl_0} \ol\cP^{0,n}_\infty\{0\}_2\llr^{\dl_1}
\ol\cP^{0,n}_\infty\{1\}_3\cdots
\label{v94}\\
&& \qquad
 \llr^{\dl_{N-1}} \ol\cP^{0,n}_\infty\{N-1\}_{N+1}
\llr^{\dl_N} \ol\cP^{0,n}_\infty\{N\}_{N+2}\llr^{\dl_N}
\ol\cP^{0,n}_\infty\{N\}_{N+3}, \nonumber
\een
of graded densities of antifield number $\leq N+3$ which is
assumed to satisfy the homology regularity condition. This
condition states that any $\dl_{k<N-1}$-cycle $\f\in
\ol\cP_\infty^{0,n}\{k\}_{k+3}\subset
\ol\cP_\infty^{0,n}\{k+1\}_{k+3}$ is a $\dl_{k+1}$-boundary. The
nilpotency property of the boundary operator $\dl_N$ (\ref{v92})
implies the NI (\ref{v64}) and the  $(k\leq N)$-stage NI
\mar{v93}\beq
\op\sum_{0\leq|\La|} \Delta_{r_k}^{r_{k-1},\La}d_\La
(\op\sum_{0\leq|\Si|} \Delta_{r_{k-1}}^{r_{k-2},\Si}\ol c_{\Si
r_{k-2}}) +  \ol\dl(\op\sum_{0\leq |\Si|,
|\Xi|}h_{r_k}^{(r_{k-2},\Si)(A,\Xi)}\ol c_{\Si r_{k-2}}\ol s_{\Xi
A})=0. \label{v93}
\eeq

\section{The inverse second Noether theorem}

Second Noether theorems in different variants relate the NI and
higher-stage NI to the gauge and higher-stage gauge symmetries of
Lagrangian theory \cite{jpa05,jmp05,fulp}. However, the notion of
gauge symmetry of Lagrangian theory meets difficulties. In
particular, it may happen that gauge symmetries are not assembled
into an algebra, or they form an algebra on-shell
\cite{fulp02,jmp08,gom,book09}. At the same time, NI are well
defined (Theorem \ref{t4}). Therefore, one can use the inverse
second Noether theorem (Theorem \ref{t5}) in order to obtain gauge
symmetries of degenerate Lagrangian theory. This theorem
associates to the antifield KT complex (\ref{v94}) the cochain
sequence (\ref{w36}) of ghosts, whose ascent operator (\ref{w108})
provides gauge and higher-stage gauge symmetries of a Lagrangian
field theory.

Given the DGA $\ol\cP^*_\infty\{N\}$ (\ref{v91}), let us consider
the DGA
\mar{w5}\beq
\cP^*_\infty\{N\}=\cP^*_\infty[V_N\times\cdots V_1\times
V,F;Y,E\times E_1\times\cdots \times E_N] \label{w5}
\eeq
possessing the local basis $\{s^A, c^r, c^{r_1}, \ldots,
c^{r_N}\}$ of Grassmann parity $[c^{r_k}]=([\ol c_{r_k}]+1){\rm
mod}\,2$ and antifield number ${\rm Ant}[c^{r_k}]=-(k+1)$. Its
elements $c^{r_k}$, $k\in\Bbb N$, are called the ghosts of ghost
number gh$[c^{r_k}]=k+1$ \cite{barn,gom}.

\begin{theo} \label{t5} \mar{t5}
Given the KT complex (\ref{v94}), the graded commutative ring
$\cP_\infty^0\{N\}$ is split into the cochain sequence
\mar{w36}\beq
0\to \cS^0_\infty[F;Y]\ar^{\bu} \cP^0_\infty\{N\}_1\ar^{\bu}
\cP^0_\infty\{N\}_2\ar^{\bu}\cdots, \label{w36}
\eeq
with the odd ascent operator
\mar{w108,33,38}\ben
&& \bu=u + \op\sum_{1\leq k\leq N} u_{(k)}, \label{w108}\\
&& u= u^A\frac{\dr}{\dr s^A}, \qquad u^A =\op\sum_{0\leq|\La|}
c^r_\La\eta(\Delta^A_r)^\La, \label{w33}\\
&& u_{(k)}= u^{r_{k-1}}\frac{\dr}{\dr c^{r_{k-1}}}, \qquad
u^{r_{k-1}}=\op\sum_{0\leq|\La|}
c^{r_k}_\La\eta(\Delta^{r_{k-1}}_{r_k})^\La, \qquad k=1,\ldots,N,
\label{w38}\\
&&\eta (f)^\La = \op\sum_{0\leq|\Si|\leq k-|\La|}(-1)^{|\Si+\La|}
C^{|\Si|}_{|\Si+\La|} d_\Si f^{\Si+\La}, \qquad
C^a_b=\frac{b!}{a!(b-a)!}. \nonumber
\een
\end{theo}

The components $u$ (\ref{w33}), $u_{(k)}$ (\ref{w38}) of the gauge
operator $\bu$ (\ref{w108}) are the above mentioned gauge and
higher-stage gauge symmetries of reducible Lagrangian theory,
respectively. Indeed, let us consider the total DGA
$P^*_\infty\{N\}$ generated by original fields, ghosts and
antifields
\mar{f10}\beq
\{s^A, c^r, c^{r_1}, \ldots, c^{r_N},\ol s_A,\ol c_r, \ol c_{r_1},
\ldots, \ol c_{r_N}\}. \label{f10}
\eeq
It contains subalgebras $\ol\cP^*_\infty\{N\}$ (\ref{v91}) and
$\cP^*_\infty\{N\}$ (\ref{w5}), whose derivations $\dl_N$
(\ref{v92}) and $\bu$ (\ref{w108}) are prolonged to
$P^*_\infty\{N\}$. Let us extend an original Lagrangian $L$ to the
Lagrangian
\mar{w8}\beq
L_e=\cL_e\om=L+L_1=L + \op\sum_{0\leq k\leq N}
c^{r_k}\Delta_{r_k}\om=L +\dl_N( \op\sum_{0\leq k\leq N}
c^{r_k}\ol c_{r_k}\om) \label{w8}
\eeq
of zero antifield number. It is readily observed that the KT
differential $\dl_N$ is a variational symmetry of the Lagrangian
$L_e$ (\ref{w8}), i.e., we have the equalities
\mar{w19,20}\ben
&& \frac{\op\dl^\lto (c^r\Delta_r)}{\dl \ol s_A}\cE_A\om
=u^A\cE_A\om=d_H\si_0, \label{w19}\\
&&  [\frac{\op\dl^\lto (c^{r_i}\Delta_{r_i})}{\dl \ol s_A}\cE_A
+\op\sum_{k<i} \frac{\op\dl^\lto (c^{r_i}\Delta_{r_i})}{\dl \ol
c_{r_k}}\Delta_{r_k}]\om= d_H\si_i, \qquad i=1,\ldots,N.
\label{w20}
\een
A glance at the equality (\ref{w19}) shows that the graded
derivation $u$ (\ref{w33}) is a variational symmetry of an
original Lagrangian $L$. Parameterized by ghosts $c^r$, it is a
gauge symmetry of $L$ \cite{jmp05,cmp04}.  The equalities
(\ref{w20}) are brought into the form
\mar{w34}\beq
\op\sum_{0\leq|\Si|} d_\Si u^{r_{i-1}}\frac{\dr}{\dr
c^{r_{i-1}}_\Si} u^{r_{i-2}} =\ol\dl(\al^{r_{i-2}}), \qquad
\al^{r_{i-2}} = -\op\sum_{0\leq|\Si|}
\eta(h_{r_i}^{(r_{i-2})(A,\Xi)})^\Si d_\Si(c^{r_i} \ol s_{\Xi A}).
\label{w34}
\eeq
It follows that graded derivations $u_{(k)}$ (\ref{w38}) are the
$k$-stage gauge symmetries of reducible Lagrangian theory
\cite{jpa05,jmp05,jmp08}.

We agree to call $\bu$ (\ref{w108}) the gauge operator. In
contrast with the KT one, this operator need not be nilpotent. We
say that gauge and higher-stage gauge symmetries of Lagrangian
theory are algebraically closed if the gauge operator $\bu$ can be
extended to nilpotent graded derivation
\mar{w109}\beq
\bb=\bu+ \xi= u^A\dr_A + \op\sum_{1\leq k\leq N}(u^{r_{k-1}}
+\xi^{r_{k-1}})\dr_{r_{k-1}}+\xi^{r_N}\dr_{r_N} \label{w109}
\eeq
of ghost number 1 where the coefficients $\xi^{r_{k-1}}$ are at
least quadratic in ghosts \cite{ijgmmp05,jmp08}. This extension is
called the BRST operator. It brings the cochain sequence
(\ref{w36}) into the BRST complex.

\section{BRST extended field theory}

Lagrangian field theory extended to ghosts and antifields
exemplifies so called field-antifield Lagrangian theories of the
following type \cite{lmp08,book09}.

Given a fibre bundle $Z\to X$ and a vector bundle $Z'\to X$, let
us consider a DGA $\cP^*_\infty[\ol Z^*,Z';Z,\ol Z'^*]$ with a
local basis $\{z^a,\ol z_a\}$, where $[\ol z_a]=([z^a]+1){\rm
mod}\,2$. One can think of its elements $z^a$ and $\ol z_a$ as
being fields and antifields, respectively. Its submodule
$\cP^{0,n}_\infty[\ol Z^*,Z';Z,\ol Z'^*]$ of horizontal densities
is provided with the binary operation
\mar{f11}\beq
\{\gL\om,\gL'\om\}=[\frac{\op\dl^\lto \gL}{\dl \ol z_a}\frac{\dl
\gL'}{\dl z^a} + (-1)^{[\gL][\gL']}\frac{\op\dl^\lto \gL'}{\dl \ol
z_a}\frac{\dl \gL}{\dl z^a}]\om, \label{f11}
\eeq
called the antibracket by analogy with that in field-antifield
BRST theory \cite{gom}. One treats this operation as {\it sui
generis} odd Poisson structure \cite{ala,barn98}. Let us associate
to a Lagrangian $\gL\om$ the odd graded derivations
\mar{w37}\beq
\up_\gL=\frac{\op\dl^\lto \gL}{\dl \ol z_a} \frac{\dr}{\dr z^a},
\qquad \ol\up_\gL=\frac{\op\dr^\lto}{\dr \ol z_a}\frac{\dl
\gL}{\dl z^a}. \label{w37}
\eeq
Then the following conditions are equivalent: (i) the graded
derivation $\up_\gL$ (\ref{w37}) is a variational symmetry of a
Lagrangian $\gL\om$,  (ii) so is the graded derivation
$\ol\up_\gL$, (iii) the relation
\mar{w44}\beq
\{\gL\om,\gL\om\}=2\frac{\op\dl^\lto \gL}{\dl \ol z_a}\frac{\dl
\gL}{\dl z^a}\om =d_H\si \label{w44}
\eeq
holds \cite{lmp08}. This relation is called the (classical) master
equation.

Let us consider an original Lagrangian $L$ and its extension $L_e$
(\ref{w8}), together with the odd graded derivations (\ref{w37})
which read
\be
\up_e= \frac{\op\dl^\lto \cL_1}{\dl \ol s_A}\frac{\dr}{\dr s^A} +
\op\sum_{0\leq k\leq N} \frac{\op\dl^\lto \cL_1}{\dl \ol
c_{r_k}}\frac{\dr}{\dr c^{r_k}}, \qquad
 \ol\up_e= \frac{\rdr }{\dr
\ol s_A}\frac{\dl\cL_1}{\dl s^A} + [\frac{\rdr }{\dr \ol
s_A}\frac{\dl\cL}{\dl s^A} +\op\sum_{0\leq k\leq N} \frac{\rdr
}{\dr \ol c_{r_k}}\frac{\dl \cL_1}{\dl c^{r_k}}].
\ee
An original Lagrangian $L$ trivially satisfies the master
equation. A goal is to extend it to a nontrivial solution
\mar{w61}\beq
L+L_1+L_2+\cdots =L_e+L' \label{w61}
\eeq
of the master equation by means of terms $L_i$ of polynomial
degree $i>1$ in ghosts and of zero antifield number. Such an
extension need not exists. One can show the following
\cite{lmp08,book09}.

\begin{theo} \label{t6} \mar{t6}
If the gauge operator $\bu$ (\ref{w108}) admits the BRST extension
$\bb$ (\ref{w109}), the Lagrangian
\mar{w133}\beq
L_E=L_e + \op\sum_{1\leq k\leq N}\xi^{r_{k-1}}\ol c_{r_{k-1}}\om=
L+\bb( \op\sum_{0\leq k\leq N} c^{r_{k-1}}\ol c_{r_{k-1}})\om
+d_H\si \label{w133}
\eeq
satisfies the master equation $\{L_E,L_E\}=0$.
\end{theo}

The proof of Theorem \ref{t6} gives something more. The gauge
operator (\ref{w108}) admits the BRST extension (\ref{w109}) only
if the higher-stage gauge symmetry conditions hold off-shell. For
instance, this is the case of irreducible and abelian reducible
Lagrangian theories. In abelian reducible theories, the gauge
operator $\bu$ itself is nilpotent. In irreducible Lagrangian
theory, the gauge operator admits a nilpotent BRST extension if
gauge transformations form a Lie algebra.

\section{Gauge theory of principal connections}

Let us consider gauge theory of principal connections on a
principal bundle $P\to X$ with a structure Lie group $G$.
Principal connections are $G$-equivariant connections on $P\to X$
and, therefore, they are represented by sections of the quotient
bundle
\mar{f30}\beq
C=J^1P/G\to X \label{f30}
\eeq
\cite{book,book00,book09}. This is an affine bundle coordinated by
$(x^\la, a^r_\la)$ such that, given a section $A$ of $C\to X$, its
components $A^r_\la=a^r_\la\circ A$ are coefficients of the
familiar local connection form (i.e., gauge potentials).
Therefore, one calls $C$ (\ref{f30}) the bundle of principal
connections. A key point is that its first order jet manifold
$J^1C$ admits the canonical splitting over $C$ given by the
coordinate expression
\mar{f31}\beq
a_{\la\m}^r = \frac12\cF^r_{\la\m} +  \frac12\cS^r_{\la\m}
=\frac{1}{2}(a_{\la\m}^r + a_{\m\la}^r
 - c_{pq}^r a_\la^p a_\m^q) + \frac{1}{2}
(a_{\la\m}^r - a_{\m\la}^r + c_{pq}^r a_\la^p a_\m^q), \label{f31}
\eeq
where $c_{pq}^r$ are the structure constants of the Lie algebra
$\ccG$ of $G$, and $F^r_{\la\m}=\cF^r_{\la\m}\circ J^1A$ is the
strength of a principal connection $A$.

There is a unique (Yang--Mills) quadratic gauge invariant
Lagrangian $L_{YM}$ on $J^1C$ which factorizes through the
component $\cF^r_{\la\m}$ of the splitting (\ref{f31}). Its gauge
symmetries are $G$-invariant vertical vector fields on $P$. They
are given by sections $\xi=\xi^re_r$ of the Lie algebra bundle
$V_GP=VP/G$, and define vector fields
\mar{f33}\beq
\xi=(-c^r_{ji}\chi^ja^i_\la + \dr_\la\chi^r)\dr^\la_r \label{f33}
\eeq
on the bundle of principal connections $C$ such that
$\bL_{J^1\chi}L_{YM}=0$. The corresponding irreducible NI read
\be
 c^r_{ji}a^i_\la\cE_r^\la + d_\la\cE_j^\la=0.
\ee

As a consequence, the basis $(a^r_\la,c^r,\ol a^\la_r,\ol c_r)$
for the BRST extended gauge theory consists of gauge potentials
$a^r_\la$, ghosts $c^r$ of ghost number 1, and antifields $\ol
a^\la_r$, $\ol c_r$ of antifield numbers 1 and 2, respectively.
Replacing gauge parameters $\chi^r$ in $\chi$ (\ref{f33}) with odd
ghost $c^r$, we obtain the gauge operator $\bu$ (\ref{w108}),
whose nilpotent extension is the well known BRST operator
\be
\bb= (-c^r_{ji}c^ja^i_\la + c^r_\la)\frac{\dr}{\dr a_\la^r}
-\frac12 c^r_{ij}c^ic^j\frac{\dr}{\dr c^r}.
\ee
Hence, the Yang--Mills Lagrangian is extended to a solution of the
master equation
\be
L_E=L_{YM}+ (-c^r_{ij}c^ja^i_\la + c^r_\la)\ol a^\la_r\om -\frac12
c^r_{ij}c^ic^j\ol c_r\om.
\ee

\section{Topological
 Chern--Simons gauge theory}

Gauge symmetries of topological Chern--Simons (henceforth CS)
gauge theory are wider than those of the Yang--Mills gauge one.

One usually considers CS theory whose Lagrangian is the local CS
form derived from the local transgression formula for the second
Chern characteristic form. The global CS Lagrangian is well
defined, but depends on a background gauge potential
\cite{bor06,bor07,fat05,mpl}.

The fibre bundle $J^1P\to C$ is a trivial $G$-principal bundle
canonically isomorphic to $C\times P\to C$. This bundle admits the
canonical principal connection
\be
\cA =dx^\la\ot(\dr_\la +a_\la^p \ve_p) + da^r_\la\ot\dr^\la_r
\ee
\cite{book00}. Its curvature defines the canonical $V_GP$-valued
2-form
\mar{f34}\beq
\gF =(da_\m^r\w dx^\m + \frac{1}{2} c_{pq}^r a_\la^p a_\m^q
dx^\la\w dx^\m)\ot e_r \label{f34}
\eeq
on $C$. Given a section $A$ of $C\to X$, the pull-back
\mar{r47'}\beq
F_A=A^*\gF=\frac12 F^r_{\la\m}dx^\la\w dx^\m\ot e_r \label{r47'}
\eeq
of $\gF$ onto $X$ is the strength form of a gauge potential $A$.

Let $I_k(e)=b_{r_1\ldots r_k}e^{r_1}\cdots e^{r_k}$ be a
$G$-invariant polynomial of degree $k>1$ on the Lie algebra
$\ccG$. With $\gF$ (\ref{f34}), one can associate to $I_k$ the
closed gauge-invariant $2k$-form
\be
P_{2k}(\gF)=b_{r_1\ldots r_k}\gF^{r_1}\w\cdots\w \gF^{r_k}
\ee
on $C$. Given a section $B$ of $C\to X$, the pull-back $
P_{2k}(F_B)=B^*P_{2k}(\gF)$ of $P_{2k}(\gF)$ is a closed
characteristic form on $X$. Let the same symbol stand for its
pull-back onto $C$. Since $C\to X$ is an affine bundle and the de
Rham cohomology of $C$ equals that of $X$, the forms $P_{2k}(\gF)$
and $P_{2k}(F_B)$ possess the same cohomology class
$[P_{2k}(\gF)]=[P_{2k}(F_B)]$ for any principal connection $B$.
Thus, $I_k(e)\mapsto [P_{2k}(F_B)]\in H^*(X)$ is the familiar Weil
homomorphism. Furthermore, we obtain the transgression formula
\mar{r65}\beq
P_{2k}(\gF)-P_{2k}(F_B)=d\gS_{2k-1}(a,B) \label{r65}
\eeq
on $C$ \cite{mpl,book09}. Its pull-back by means of a section $A$
of $C\to X$ gives the transgression formula
\be
P_{2k}(F_A)-P_{2k}(F_B)=d \gS_{2k-1}(A,B)
\ee
on $X$. For instance, if $P_{2k}(\gF)$ is the characteristic Chern
$2k$-form, then $\gS_{2k-1}(a,B)$ is the CS $(2k-1)$-form.

In particular, one can choose the local section $B=0$. Then,
$\gS_{2k-1}(a,0)$ is the local CS form. Let $\gS_{2k-1}(A,0)$ be
its pull-back onto $X$ by means of a section $A$ of $C\to X$. Then
the CS form $\gS_{2k-1}(a,B)$ (\ref{r65}) admits the decomposition
\mar{r75}\beq
\gS_{2k-1}(a,B)=\gS_{2k-1}(a,0) -\gS_{2k-1}(B,0) +dK_{2k-1}.
\label{r75}
\eeq
The transgression formula (\ref{r65}) also yields the
transgression formula
\mar{0742}\ben
&& P_{2k}(\cF)-P_{2k}(F_B)=d_H(h_0 \gS_{2k-1}(a,B)), \nonumber\\
&& h_0 \gS_{2k-1}(a,B)=k\op\int^1_0 \cP_{2k}(t,B)dt, \label{0742}\\
&& \cP_{2k}(t,B)=b_{r_1\ldots
r_k}(a^{r_1}_{\m_1}-B^{r_1}_{\m_1})dx^{\m_1}\w
\cF^{r_2}(t,B)\w\cdots \w \cF^{r_k}(t,B),\nonumber\\
&& \cF^{r_j}(t,B)= \frac12[ ta^{r_j}_{\la_j\m_j}
+(1-t)\dr_{\la_j}B^{r_j}_{\m_j}
- ta^{r_j}_{\m_j\la_j} -(1-t)\dr_{\m_j}B^{r_j}_{\la_j}+\nonumber\\
&& \qquad \frac12c^{r_j}_{pq} (ta^p_{\la_j}
+(1-t)B^p_{\la_j})(ta^q_{\m_j} +(1-t)B^q_{\m_j}]dx^{\la_j}\w
dx^{\m_j}\ot e_r,\nonumber
\een
on $J^1C$. If $2k-1=\di X$, the density
$L_{CS}(B)=h_0\gS_{2k-1}(a,B)$ (\ref{0742}) is the global CS
Lagrangian of topological CS theory. The decomposition (\ref{r75})
induces the decomposition
\mar{0747}\beq
L_{CS}(B)=h_0\gS_{2k-1}(a,0) -\gS_{2k-1}(B,0) +d_H h_0K_{2k-1}.
\label{0747}
\eeq

For instance, if $\di X=3$, the global CS Lagrangians reads
\be
&& L_{CS}(B)= [\frac12 h_{mn} \ve^{\al\bt\g}a^m_\al(\cF^n_{\bt\g}
-\frac13 c^n_{pq}a^p_\bt a^q_\g)]\om -  \\
&& \qquad [\frac12 h_{mn} \ve^{\al\bt\g}B^m_\al(F(B)^n_{\bt\g}
-\frac13 c^n_{pq}B^p_\bt B^q_\g)]\om -d_\al(h_{mn}
\ve^{\al\bt\g}a^m_\bt B^n_\g)\om,
\ee
where $\ve^{\al\bt\g}$ is the skew-symmetric Levi--Civita tensor.

Since the density $-\gS_{2k-1}(B,0) +d_Hh_0K_{2k-1}$ is
variationally trivial, the global CS Lagrangian (\ref{0747})
possesses the same NI and gauge symmetries as the local one
$L_{CS}=h_0\gS_{2k-1}(a,0)$. They are the following.

In contrast with the Yang--Mills Lagrangian, the CS one
$L_{CS}(B)$ is independent of a metric on $X$. Therefore, its
gauge symmetries are all $G$-invariant vector fields on a
principal bundle $P$. They are identified to sections
\be
v_P=\tau^\la\dr_\la +\chi^r e_r
\ee
of the vector bundle $T_GP=TP/G\to X$, and yield vector fields
\mar{0653}\beq
v_C=\tau^\la\dr_\la +(-c^r_{pq}\chi^pa^q_\la +\dr_\la \chi^r
-a^r_\m\dr_\la \tau^\m)\dr^\la_r \label{0653}
\eeq
on the bundle of principal connections $C$ \cite{book00,book09}.
One can show that they are variational and, consequently, gauge
symmetries of the global CS Lagrangian $L_{CS}(B)$. The vertical
part
\mar{0785}\beq
v_V=(-c^r_{pq}\chi^pa^q_\la +\dr_\la \chi^r -a^r_\m\dr_\la \tau^\m
-\tau^\m a^r_{\m\la} )\dr^\la_r \label{0785}
\eeq
of vector fields $v_C$ (\ref{0653}) is also a variational symmetry
of $L_{CS}(B)$.

As a consequence, the basis $(a^r_\la, c^\la, c^r, \ol a_r^\la,
\ol c_\la, \ol c_r)$ of BRST extended CS theory consists of even
fields $a^r_\la$, ghosts $c^\la$, $c^r$ and antifields $\ol
a_r^\la$, $\ol c_\la$, $\ol c_r$. Substituting the ghosts $c^\la$,
$c^r$ for gauge parameters in the vector field $v_V$ (\ref{0785}),
we obtain the gauge operator
\mar{0781}\beq
\bu=(-c^r_{pq}c^pa^q_\la + c^r_\la -c^\m_\la a^r_\m -c^\m
a_{\m\la}^r)\dr^\la_r. \label{0781}
\eeq
The corresponding irreducible NI read
\be
 -c^r_{ji}a^i_\la\cE_r^\la - d_\la\cE_j^\la=0, \qquad -
 a^r_{\m\la}\cE^\la_r +d_\la(a^r_\m\cE^\la_r)=0.
\ee

The gauge operator (\ref{0781}) admits the nilpotent BRST
extension
\be
\bb=(-c^r_{ji}c^ja^i_\la + c^r_\la -c^\m_\la a^r_\m -c^\m
a_{\m\la}^r)\frac{\dr}{\dr a_\la^r} -\frac12
c^r_{ij}c^ic^j\frac{\dr}{\dr c^r} +c^\la_\m c^\m\frac{\dr}{\dr
c^\la}.
\ee
Accordingly, the CS Lagrangian  is extended to the proper solution
of the master equation
\be
L_E=L_{CS}+ [(-c^r_{ij}c^ja^i_\la + c^r_\la)\ol a^\la_r -\frac12
c^r_{ij}c^ic^j\ol c_r + c^\la_\m c^\m \ol c_\la]\om.
\ee

\section{Field theory on composite bundles}

Let us consider a composite fibre bundle
\mar{1.34}\beq
Y\to \Si\to X, \label{1.34}
\eeq
where $\pi_{Y\Si}: Y\to\Si$ and $\pi_{\Si X}: \Si\to X$ are fibre
bundles. It is provided with fibred coordinates
$(x^\la,\si^m,y^i)$, where $(x^\m,\si^m)$ are bundle coordinates
on $\Si\to X$, i.e., the transition functions of coordinates
$\si^m$ are independent of coordinates $y^i$. The following facts
make composite bundles useful for physical applications
\cite{book,sau,book09}.

Given a composite bundle (\ref{1.34}), let $h$ be a global section
of $\Si\to X$. Then the restriction
\mar{S10}\beq
Y_h=h^*Y \label{S10}
\eeq
of the fibre bundle $Y\to\Si$ to $h(X)\subset \Si$ is a subbundle
of the fibre bundle $Y\to X$.

Every section $s$ of the fibre bundle $Y\to X$ is a composition of
the section $h=\pi_{Y\Si}\circ s$ of the fibre bundle $\Si\to X$
and some section of the fibre bundle $Y\to \Si$ over $h(X)\subset
\Si$.

Let $J^1\Si$, $J^1_\Si Y$, and $J^1Y$ be jet manifolds of the
fibre bundles $\Si\to X$, $Y\to \Si$ and $Y\to X$, respectively.
They are provided with the adapted coordinates $(x^\la ,\si^m,
\si^m_\la)$, $( x^\la ,\si^m, y^i, \wt y^i_\la, y^i_m)$ and
$(x^\la ,\si^m, y^i, \si^m_\la ,y^i_\la)$. There is the canonical
map
\be
\vr : J^1\Si\op\times_\Si J^1_\Si Y\ar_Y J^1Y, \qquad
y^i_\la\circ\vr=y^i_m{\si}^m_{\la} +\wt y^i_{\la}.
\ee
Due to this map, any pair of connections
\mar{b1.113}\ben
&& A_\Si=dx^\la\ot (\dr_\la + A_\la^i\dr_i) +d\si^m\ot (\dr_m +
A_m^i\dr_i), \label{b1.113} \\
&& \G=dx^\la\ot (\dr_\la + \G_\la^m\dr_m) \nonumber
\een
on fibre bundles $Y\to \Si$ and $\Si\to X$, respectively,  yields
the composite connection
\mar{b1.114}\beq
\g=A_\Si\circ\G=dx^\la\ot (\dr_\la +\G_\la^m\dr_m + (A_\la^i +
A_m^i\G_\la^m)\dr_i) \label{b1.114}
\eeq
on the fibre bundle $Y\to X$. For instance, let us consider a
vector field $\tau$ on the base $X$, its horizontal lift $\G\tau$
onto $\Si$ by means of the connection $\G$ and, in turn, the
horizontal lift $A_\Si(\G\tau)$ of $\G\tau$ onto $Y$ by means of
the connection $A_\Si$. Then $A_\Si(\G\tau)$ is the horizontal
lift of $\tau$ onto $Y$ by means of the composite connection $\g$
(\ref{b1.114}).

Given a composite bundle $Y$ (\ref{1.34}), there is the exact
sequence of bundles
\mar{63}\beq
0\to V_\Si Y\to VY\to Y\op\times_\Si V\Si\to 0, \label{63}
\eeq
where $V_\Si Y$ is the vertical tangent bundle of the fibre bundle
$Y\to\Si$. Every connection $A$ (\ref{b1.113}) on the fibre bundle
$Y\to\Si$ yields the splitting
\be
\dot y^i\dr_i + \dot\si^m\dr_m= (\dot y^i -A^i_m\dot\si^m)\dr_i +
\dot\si^m(\dr_m+A^i_m\dr_i)
\ee
of the exact sequences (\ref{63}). This splitting defines the
first order differential operator
\mar{7.10}\beq
\wt D= dx^\la\otimes(y^i_\la- A^i_\la -A^i_m\si^m_\la)\dr_i
\label{7.10}
\eeq
on the composite bundle $Y\to X$. This operator, called the
vertical covariant differential, possesses the following important
property. Let $h$ be a section of the fibre bundle $\Si\to X$ and
$Y_h$ the subbundle (\ref{S10}) of the composite bundle $Y\to X$.
Then the restriction of the vertical covariant differential $\wt
D$ (\ref{7.10}) to $J^1Y_h\subset J^1Y$ coincides with the
familiar covariant differential relative to the pull-back
connection
\mar{mos83}\beq
A_h=h^*A_\Si=dx^\la\ot[\dr_\la+((A^i_m\circ h)\dr_\la h^m +(A\circ
h)^i_\la)\dr_i] \label{mos83}
\eeq
on $Y_h\to X$  \cite{book,book00,book09}.

The peculiarity of field theory on a composite bundle (\ref{1.34})
is that its Lagrangian depends on a connection on $Y\to\Si$, but
not $Y\to X$, and it factorizes through the vertical covariant
differential (\ref{7.10}). This is the case of field theories with
broken symmetries, spinor fields, gauge gravitation theory
\cite{book00,sard98a,sard02,sard06,sard06a,book09}.

\section{Symmetry
 breaking and Higgs fields}

In gauge theory on a principal bundle $P\to X$, a symmetry
breaking is defined as reduction of the structure Lie group $G$ of
this principal bundle to a closed (consequently, Lie) subgroup $H$
of exact symmetries
\cite{castr,book,keyl,nik,sard92,sard06a,book09}.

By virtue of the well-known theorem,  reduction of the structure
group of a principal bundle takes place iff there exists a global
section $h$ of the quotient bundle $P/H\to X$. This section  is
treated as a Higgs field. Thus, we have the composite bundle
\mar{b3223a}\beq
P\to P/H\to X, \label{b3223a}
\eeq
where $P\to P/H$ is a principal bundle with the structure group
$H$ and  $\Si=P/H\to X$ is a $P$-associated fibre bundle with the
typical fibre $G/H$. Moreover, there is one-to-one correspondence
between the global sections $h$ of $\Si\to X$ and reduced
$H$-principal subbundles $P^h=\pi_{P\Si}^{-1}(h(X))$ of $P$.

Let $Y\to \Si$ be a vector bundle associated to the $H$-principal
bundle $P\to \Si$. Then sections of the composite bundle $Y\to
\Si\to X$ describe matter fields with the exact symmetry group $H$
in the presence of Higgs fields. Given bundle coordinates
$(x^\la,\si^m,y^i)$ on $Y$, these sections are locally represented
by  pairs $(\si^m(x), y^i(x))$. Given a global section $h$ of
$\Si\to X$, sections of the vector bundle $Y_h$ (\ref{S10})
describe matter fields in the presence of the background Higgs
field $h$. Moreover, for different Higgs fields $h$ and $h'$, the
fibre bundles $Y_h$ and $Y_{h'}$ need not be equivalent
\cite{book,sard92,sard06a}.

Note that $Y\to X$ fails to be associated to a principal bundle
$P\to X$ with the structure group $G$ and, consequently, it need
not admit a principal connection. Therefore, one should consider a
principal connection (\ref{b1.113}) on the fibre bundle $Y\to
\Si$, and a Lagrangian on $J^1Y$ factorizes through the vertical
covariant differential $\wt D$ (\ref{7.10}). In the presence of a
background Higgs field $h$, the restriction of $\wt D$ to $J^1Y_h$
coincides with the covariant differential relative to the
pull-back connection (\ref{mos83}) on $Y_h\to X$.

Riemannian and pseudo-Riemannian metrics on a manifold $X$
exemplify classical Higgs fields. Let $X$ be an oriented
four-dimensional smooth manifold and $LX$ the fibre bundle of
linear frames in the tangent spaces to $X$. It is a principal
bundle with the structure group $GL_4=GL^+(4,\Bbb R)$. This
structure group is always reducible to its maximal compact
subgroup $SO(4)$. The corresponding global sections of the
quotient bundle $LX/SO(4)$ are Riemannian metrics on $X$. However,
the reduction of the structure group $GL_4$ of $LX$ to its Lorentz
subgroup $SO(1,3)$ and a pseudo-Riemannian metric on $X$ need not
exist.

Note that, if $G=GL_4$ and $H=SO(1,3)$, we are in the case of so
called reductive $G$-structure \cite{godina03} when the Lie
algebra $\ccG$ of $G$ is the direct sum
\mar{g13}\beq
{\cal G} = {\got h} \oplus {\got m} \label{g13}
\eeq
of the Lie algebra ${\got h}$ of $H$ and a subspace ${\got
m}\subset \ccG$ such that $ad(g)({\got m})\subset {\got m}$, $g\in
H$. In this case, the pull-back of the ${\got h}$-valued component
of any principal connection on $P$ onto a reduced subbundle $P^h$
is a principal connection on $P^h$.

\section{Natural and gauge-natural bundles}

A connection $\G$ on a fibre bundle $Y\to X$ defines the
horizontal lift $\G\tau$ onto $Y$ of any vector field $\tau$ on
$X$. There is the category of natural bundles \cite{kol,terng}
which admit the functorial lift $\wt\tau$ onto $T$ of any vector
field $\tau$ on $X$ such that $\tau\mapsto\ol\tau$ is a
monomorphism of the Lie algebra of vector field on $X$ to that on
$T$. One can think of the lift $\wt\tau$ as being an infinitesimal
generator of a local one-parameter group of general covariant
transformations of $T$. The corresponding Noether current
$\gJ_{\wt\tau}$ is the energy-momentum flow along $\tau$
\cite{book,sard97b,sard98a,book09}.

Natural bundles are exemplified by tensor bundles over $X$.
Moreover, all bundles associated to the principal frame bundle
$LX$ are natural bundles. The bundle
\mar{gr14}\beq
C_K=J^1LX/GL_4 \label{gr14}
\eeq
of principal connections on $LX$ is not associated to $LX$, but it
is also a natural bundle \cite{book,book00}.

Note that a spinor bundle $S^g$ associated to a pseudo-Riemannian
metric $g$ on $X$ admits the canonical lift of any vector field on
$X$ onto $S^h$. It is called Kosmann's Lie derivative
\cite{fat98,god}. Such a lift is a property of any reductive
$G$-structure \cite{godina03}. However, this lift fails to be
functorial, and spinor bundles are not natural.

In a more general setting, higher order natural bundles and
gauge-natural bundles are called into play
\cite{eck,fat03,kol,terng}. Note that the linear frame bundle $LX$
over a manifold $X$ is the set of first order jets of local
diffeomorphisms of $\Bbb R^n$ to $X$, $n=\di X$, at the origin of
$\Bbb R^n$. Accordingly, one considers $r$-order frame bundles
$L^rX$ of $r$-order jets of local diffeomorphisms of $\Bbb R^n$ to
$X$. Furthermore, given a principal bundle $P\to X$ with a
structure group $G$, the $r$-order jet bundle $J^1P\to X$ of its
sections fails to be a principal bundle. However, the product
$W^rP=L^rX\times J^rP$ is a principal bundle with the structure
group $W^r_nG$ which is a semi direct product of the group $G^r_n$
of invertible $r$-order jets of maps $\Bbb R^n$ to itself at its
origin (e.g., $G^1_n=GL(n,\Bbb R)$) and the group $T^r_nG$ of
$r$-order jets of morphisms $\Bbb R^n\to G$ at the origin of $\Bbb
R^n$. Moreover, if $Y\to X$ is a fibre bundle associated to $P$,
the jet bundle $J^rY\to X$ is a vector bundle associated to the
principal bundle $W^rP$. It exemplifies gauge natural bundles,
which can described as fibre bundles associated to principal
bundles $W^rP$. Natural bundles are gauge natural bundles for a
trivial $G=1$. The bundle of principal connections $C$ (\ref{f30})
is a first order gauge natural bundle. This fact motivates
somebody to develop generalized gauge theory on gauge natural
bundles \cite{bib,castro,fat03,fat04}.

\section{Gauge gravitation theory}

Gauge gravitation theory (see
\cite{ald08,ali,chams,heh,iva,klaud,obukh,sard06,vign} for a
survey) can be described as a field theory on natural bundles over
an oriented four-dimensional manifold $X$ whose dynamic variables
are linear connections and pseudo-Riemannian metrics on $X$
\cite{ijgmmp05,book00,sard02,sard06,book09}.

Linear connections on $X$ (henceforth world connection) are
principal connections on the linear frame bundle $LX$  of $X$.
They are represented by sections of the bundle of linear
connections $C_K$ (\ref{gr14}). This is provided with bundle
coordinates $(x^\la,k_\la{}^\nu{}_\al)$ such that components
$k_\la{}^\nu{}_\al\circ K=K_\la{}^\nu{}_\al$ of a section $K$ of
$C_K\to X$ are coefficient of the linear connection
\be
K=dx^\la\ot (\dr_\la + K_\la{}^\m{}_\nu \dot x^\nu\dot\dr_\mu)
\ee
on $TX$ with respect to the holonomic bundle coordinates
$(x^\la,\dot x^\la)$.

In order to describe gravity, let us assume that the linear frame
bundle $LX$ admits a Lorentz structure, i.e., reduced principal
subbundles with the structure Lorentz group. Global sections of
the corresponding quotient bundle
\mar{b3203}\beq
\Si_{\rm PR}= LX/SO(1,3)\to X \label{b3203}
\eeq
are pseudo-Riemannian (henceforth world) metrics on $X$. This fact
motivates us to treat a metric gravitational field as a Higgs
field \cite{iva,sard02,sard06}.

The total configuration space of gauge gravitation theory in the
absence of matter fields is the bundle product $\Si_{PR}\times
C_K$ coordinated by $(\si^{\al\bt},  k_\mu{}^\al{}_\bt)$. This is
a natural bundle admitting the functorial lift
\mar{gr3}\ben
&& \wt\tau_{K\Si}=\tau^\m\dr_\m +(\si^{\nu\bt}\dr_\nu \tau^\al
+\si^{\al\nu}\dr_\nu \tau^\bt)\frac{\dr}{\dr \si^{\al\bt}} +
\label{gr3}\\
&& \qquad (\dr_\nu \tau^\al k_\m{}^\nu{}_\bt -\dr_\bt \tau^\nu
k_\m{}^\al{}_\nu -\dr_\mu \tau^\nu k_\nu{}^\al{}_\bt
+\dr_{\m\bt}\tau^\al)\frac{\dr}{\dr k_\mu{}^\al{}_\bt} \nonumber
\een
of vector fields $\tau$ on $X$ \cite{ijgmmp05,book00,book09}.
These lifts are generators of one-dimensional groups of general
covariant transformations, whose gauge parameters are vector
fields on $X$.

We do not specify a gravitation Lagrangian $L_G$ on the jet
manifold $J^1(\Si_{PR}\times C_K)$, but assume that vector fields
(\ref{gr3}) exhaust its gauge symmetries. Then the Euler--Lagrange
operator $(\cE_{\al\bt} d\si^{\al\bt} + \cE^\m{}_\al{}^\bt
dk_\m{}^\al{}_\bt)\w\om$ of this Lagrangian obeys irreducible
Noether identities
\be
&&-(\si^{\al\bt}_\la +2\si^{\nu\bt}_\nu\dl^\al_\la)\cE_{\al\bt}
-2\si^{\nu\bt}d_\nu\cE_{\la\bt} +(-k_{\la\m}{}^\al{}_\bt
-k_{\nu\m}{}^\nu{}_\bt\dl^\al_\la + k_{\bt\m}{}^\al{}_\la +
k_{\m\la}{}^\al{}_\bt)\cE^\m{}_\al{}^\bt +\\
&& \qquad (-k_\m{}^\nu{}_\bt\dl^\al_\la
+k_\m{}^\al{}_\la\dl^\nu_\bt
+k_\la{}^\al{}_\bt\dl^\nu_\m)d_\nu\cE^\m{}_\al{}^\bt + d_{\m\bt}
\cE^\m{}_\la{}^\bt=0
\ee
\cite{ijgmmp05}. Taking the vertical part of vector fields
$\wt\tau_{K\Si}$ and replacing gauge parameters $\tau^\la$ with
ghosts $c^\la$, we obtain the gauge operator and its nilpotent
BRST prolongation
\be
&&u_E=u^{\al\bt}\frac{\dr}{\dr\si^{\al\bt}} +u_\m{}^\al{}_\bt
\frac{\dr}{\dr k_\mu{}^\al{}_\bt} +u^\la \frac{\dr}{\dr
c^\la}=(\si^{\nu\bt} c_\nu^\al +\si^{\al\nu}
c_\nu^\bt-c^\la\si_\la^{\al\bt})\frac{\dr}{\dr \si^{\al\bt}}+
\\
&& \qquad (c_\nu^\al k_\m{}^\nu{}_\bt -c_\bt^\nu k_\m{}^\al{}_\nu
-c_\mu^\nu k_\nu{}^\al{}_\bt +c_{\m\bt}^\al-c^\la
k_{\la\mu}{}^\al{}_\bt)\frac{\dr}{\dr k_\mu{}^\al{}_\bt} +
c^\la_\m c^\m\frac{\dr}{\dr c^\la},
\ee
but this differs from that in \cite{gron}. Accordingly, an
original Lagrangian $L_G$ is extended to a solution of the master
equation
\be
L_E= L_G + u^{\al\bt}\ol\si_{\al\bt}\om + u_\m{}^\al{}_\bt \ol
k^\m{}_\al{}^\bt\om + u^\la \ol c_\la\om,
\ee
where $\ol\si_{\al\bt}$, $\ol k^\m{}_\al{}^\bt$ and $\ol c_\la$
are corresponding antifields.

\section{Dirac spinor fields}

Dirac spinors as like as other ones are described in the Clifford
algebra terms \cite{fried,law}. The Dirac spinor structure on a
four-dimensional manifold $X$ is defined as a pair $(P^h, z_s)$ of
a principal bundle $P^h\to X$ with the structure spin group
$L_s=SL(2,\Bbb C)$ and its bundle morphism $z_s: P^h \to LX$ to
the frame bundle $LX$ \cite{avis,law}. Any such morphism
factorizes
\mar{g10}\beq
P^h \to L^hX\to LX \label{g10}
\eeq
through some reduced principal subbundle $L^hX\subset LX$ with the
structure proper Lorentz group L$=SO^\uparrow(1,3)$, whose
universal two-fold covering is $\rL_s$. The corresponding quotient
bundle $\Si_{\rm T}=LX/\rL$ is a two-fold covering of the bundle
$\Si_{\rm PR}$ (\ref{b3203}). Its global section, called a tetrad
field, defines a principal Lorentz subbundle $L^hX$ of $LX$. It
can be represented by a family of local sections $\{h_a\}_\iota$
of $LX$ on trivialization domains $U_\iota$ which take values in
$L^hX$ and possess Lorentz transition functions. They define an
atlas $\Psi^h=\{(\{h_a\}_\iota, U_\iota)\}$ of $LX$ with Lorentz
transition functions such that the corresponding pseudo-Riemannian
metric on $X$ reads $g_{\m\nu}=h^a_\m h^b_\nu\eta_{ab}$, where
$\eta_{ab}$ is the Minkowski metric.

Thus, any Dirac spinor structure is associated to a Lorentz
reduced structure, but the converse need not be true. There is the
well-known topological obstruction to the existence of a Dirac
spinor structure. For instance, a Dirac spinor structure on a
non-compact manifold $X$ exists iff $X$ is parallelizable.

Given a Dirac spinor structure (\ref{g10}), the associated Dirac
spinor bundle $S^h$ can be seen as a subbundle of the bundle of
Clifford algebras generated by the Lorentz frames $\{t_a\}\in
L^hX$ \cite{benn,law}. This fact enables one to define the
Clifford representation
\mar{g11}\beq
\g_h(dx^\m)=h^\m_a\g^a \label{g11}
\eeq
of coframes $dx^\m$ in the cotangent bundle $T^*X$ by Dirac's
matrices, and introduce the Dirac operator on $S^h$ with respect
to a principal connection on $P^h$. Then sections of a spinor
bundle $S^h$ describe Dirac spinor fields in the presence of a
tetrad field $h$. However, the representations (\ref{g11}) for
different tetrad fields fail to be equivalent. Therefore, one
meets a problem of describing Dirac spinor fields in the presence
of different tetrad fields and under general covariant
transformations.

In order to solve this problem, let us consider the universal
two-fold covering $\wt{GL}_4$ of the group $GL_4$ and the
$\wt{GL}_4$-principal bundle $\wt{LX}\to X$ which is the two-fold
covering bundle  of the frame bundle $LX$ {\cite{dabr,law,swit}.
Then we have the commutative diagram
\be
\begin{array}{ccc}
 \wt{LX} & \ar^\zeta & LX \\
 \put(0,-10){\vector(0,1){20}} &
& \put(0,-10){\vector(0,1){20}}  \\
P^h & \ar & L^hX
\end{array}
\ee
for any Dirac spinor structure (\ref{g10})
\cite{fulp94,sard98a,sard02}. As a consequence,
$\wt{LX}/\rL_s=LX/L=\Si_{\rm T}$. Since $\wt{LX}\to \Si_T$ is an
$\rL_s$-principal bundle, one can consider the associated spinor
bundle $S\to \Si_T$ whose typical fibre is a Dirac spinor space
$V_s$ \cite{book00,sard98a,sard02,book09}. We agree to call it the
universal spinor bundle because, given a tetrad field $h$, the
pull-back $S^h=h^*S\to X$ of $S$ onto $X$ is a spinor bundle on
$X$ which is associated to the $\rL_s$-principal bundle $P^h$. The
universal spinor bundle $S$ is endowed with bundle coordinates
$(x^\la, \si^\m_a,y^A)$, where $(x^\la, \si^\m_a)$ are bundle
coordinates on $\Si_T$ and $y^A$ are coordinates on the spinor
space $V_s$. The universal spinor bundle $S\to\Si_T$ is a
subbundle of the bundle of Clifford algebras which is generated by
the bundle of Minkowski spaces associated to the L-principal
bundle $LX\to\Si_T$. As a consequence, there is the Clifford
representation
\mar{L7}\beq
\g_\Si: T^*X\op\ot_{\Si_T} S \to S, \qquad \g_\Si (dx^\la)
=\si^\la_a\g^a, \label{L7}
\eeq
whose restriction to the subbundle $S^h\subset S$ restarts the
representation (\ref{g11}).

Sections of the composite bundle $S\to \Si_{\rm T}\to X$ describe
Dirac spinor fields in the presence of different tetrad fields as
follows \cite{sard98a,sard02}. Due to the splitting (\ref{g13}),
any general linear connection $K$ on $X$ (i.e., a principal
connection on $LX$) yields the connection
\mar{b3266}\ben
&& A_\Si = dx^\la\ot(\dr_\la - \frac14
(\eta^{kb}\si^a_\m-\eta^{ka}\si^b_\m)
 \si^\nu_k K_\la{}^\m{}_\nu L_{ab}{}^A{}_By^B\dr_A) +
 \label{b3266}\\
&& \qquad d\si^\m_k\ot(\dr^k_\m + \frac14 (\eta^{kb}\si^a_\m
-\eta^{ka}\si^b_\m) L_{ab}{}^A{}_By^B\dr_A) \nonumber
\een
on the universal spinor bundle $S\to\Si_T$. Its restriction to
$S^h$ is the familiar spin connection
\mar{b3212}\beq
K_h=dx^\la\ot[\dr_\la +\frac14
(\eta^{kb}h^a_\m-\eta^{ka}h^b_\m)(\dr_\la h^\m_k - h^\nu_k
K_\la{}^\m{}_\nu)L_{ab}{}^A{}_B y^B\dr_A], \label{b3212}
\eeq
defined by $K$ \cite{pon,sard97b}. The connection (\ref{b3266})
yields the vertical covariant differential
\mar{7.10'}\beq
\wt D= dx^\la\ot[y^A_\la- \frac14(\eta^{kb}\si^a_\m
-\eta^{ka}\si^b_\m)(\si^\m_{\la k} -\si^\nu_k
K_\la{}^\m{}_\nu)L_{ab}{}^A{}_By^B]\dr_A, \label{7.10'}
\eeq
on the fibre bundle $S\to X$. Its restriction to $J^1S^h\subset
J^1S$ recovers the familiar covariant differential on the spinor
bundle $S^h\to X$ relative to the spin connection (\ref{b3212}).
Combining (\ref{L7}) and (\ref{7.10'}) gives the first order
differential operator
\be
\cD=\si^\la_a\g^{aB}{}_A[y^A_\la- \frac14(\eta^{kb}\si^a_\m
-\eta^{ka}\si^b_\m)(\si^\m_{\la k} -\si^\nu_k
K_\la{}^\m{}_\nu)L_{ab}{}^A{}_By^B],
\ee
on the fibre bundle $S\to X$. Its restriction to $J^1S^h\subset
J^1S$ is the familiar Dirac operator on the spinor bundle $S^h$ in
the presence of a background tetrad field $h$ and a general linear
connection $K$.

\end{document}